\documentclass[prb,
twocolumn,
superscriptaddress,showpacs,amsmath,amssymb]{revtex4}

\usepackage{graphicx}
\usepackage{color}

\newcommand{\abs}[1]{\left \vert #1\right \vert}
\newcommand{\im}{i}
\newcommand{\unitvec}[1]{\hat{\vek{#1}}}
\newcommand{\vek}[1]{\mathbf{#1}}
\newcommand{\doo}{\partial}

\begin{document}

\title{On thermal Nieh-Yan anomaly in Weyl superfluids}

\date{\today}

\author{J. Nissinen}

\affiliation{Low Temperature Laboratory, Department of Applied Physics, Aalto University, FI-00076 AALTO, Finland}

\author{G.E.~Volovik}
\affiliation{Low Temperature Laboratory, Department of Applied Physics, Aalto University, FI-00076 AALTO, Finland}
\affiliation{Landau Institute for Theoretical Physics, 142432 Chernogolovka, Russia.}

\begin{abstract}
{We discuss the possibility of the  torsional Nieh-Yan anomaly of the type $\doo_\mu (ej^\mu_5) =\gamma T^2({\cal T}^a \wedge {\cal T}_a)$ in Weyl superfluids, where 
 $T$ is temperature and ${\cal T}^a$ is the effective or emergent torsion  from the superfluid order parameter.
As distinct from the dimensionful ultraviolet (UV) parameter $\Lambda^2$ in the conventional torsional Nieh-Yan anomaly, the parameter  $\gamma$ is dimensionless in canonical units. This suggests that such dimensionless parameter  may be fundamental, being determined by the geometry, topology and number of  chiral quantum fields in the system. By comparing this to a Weyl superfluid with low-temperature corrections, $T\ll\Delta_0$, we show that such a term does exist in the hydrodynamics of a chiral $p$-wave superfluid, such as $^3$He-A, or a chiral superconductor. We also discuss and show how other $T^2$ terms of similar form and of the same order in gradients, coming from e.g. Fermi-liquid corrections and the chiral chemical potential, can also be expressed in terms of dimensionless fundamental parameters with emergent low-energy relativistic fields. Lastly, we discuss our results in comparison to relativistic Weyl fermions and the connection of the torsional gravitational anomalies to thermal transport in Weyl systems.
}
\end{abstract}

\maketitle

\section{Introduction}

In non-relativistic topological matter, effectively quasi-relativistic description of low-energy quasiparticles with linear spectrum phenomena may emerge \cite{Volovik2003, Horava05}. In particular, in three spatial dimensions at a generic (two-fold) degenerate fermion band crossing at momentum $\mathbf{p}_W$, the Hamiltonian is of the Weyl form \cite{Herring1937, Abrikosov1971, NielsenNinomiya83, Volovik2003}
\begin{align}
H_{W} = \sigma^a e^{i}_a (p- p_W)_i + \cdots \label{eq:WeylHamiltonian}
\end{align}
where the $e^{i}_a = \partial_{p_i} H(\mathbf{p})\vert_{p_W}$ are the linear coefficients of the Hermitean Pauli matrices $\sigma^a$, $a=1,2,3$, close to the Weyl node(s) at $\mathbf{p}_W$. The net chirality {$\sum_{\{\mathbf{p}_W\}} \textrm{sgn}(\det e^{i}_a)$ vanishes \cite{NielsenNinomiya83}. For slowly varying parameters in the operator $\sigma^ae^{\mu}_a \im \doo_{\mu}\equiv i\partial_t -H_W$, the semi-classical fields $e^{\mu}_a(x) = \{e^0_{a}, e^i_a\}$ are promoted to background spacetime tetrad fields with dimensions of unity for temporal indices $e_a^0$ and velocity for spatial $e^i_a$. At the level of the Hamiltonian, the shift of the Weyl node $\mathbf{p}_W$ acts as an emergent (axial) gauge field with emergent Lorentz symmetry to the linear order but the linear expansion \eqref{eq:WeylHamiltonian} at $\vek{p}_W$ is valid at much lower scales, however. If the fermions are charged, the fermions can in addition couple to the electromagnetic vector potential via minimal coupling.

These background fields imply the chiral anomaly for the low-energy massless quasiparticles. For the applications of the chiral anomaly in Weyl semimetal and Weyl superfluids/superconductors, see e.g. \cite{NielsenNinomiya83, Volovik1986a, Volovik2003, ZyuzinBurkov12}.  In particular, the non-trivial coordinate dependence (torsion) related to the tetrads $e_a^{\mu}(x)$ in \eqref{eq:WeylHamiltonian} can lead to the gravitational Nieh-Yan (NY) anomaly 
\cite{NiehYan1982a,NiehYan1982b,Nieh2007,Yajima96,ChandiaZanelli97a,ChandiaZanelli97b,ChandiaZanelli98, Parrikar2014, FerreirosEtAl19,Nissinen2019}. Nevertheless, the NY contribution to the anomaly has remained contentious and subtle due to presence of a dimensionful ultra-violet (UV) scale $\Lambda^2$ with canonical dimensions of momentum, as required by the canonical dimensions of $e^a_{\mu}$. See, however, Ref. \onlinecite{Nissinen2019} and the discussion below. 

Here we discuss the temperature corrections to the gravitational NY anomaly and different finite temperature terms in the hydrodynamic momentum transport of chiral Weyl superfluids. The results apply for the chiral superconductors as well, when the electromagnetic potential is added $\mathbf{v}_s \to \vek{v}_s-e\vek{A}/m$ in some convenient gauge. We show that these corrections to the momentum currents include, among various other similar terms, a term originating from the torsional Nieh-Yan anomaly. For all such low-energy temperature corrections proportional to $T^2$ in the free energy, the prefactors in the corresponding terms are dimensionless, i.e. do not depend on the details of the microscopic physics, but are fully determined by geometry, topology and the number of effective fermionic species \cite{VolovikZelnikov2003,Volovik2003}. Specifically, we compare the finite temperature corrections to the lowest order gradient terms in free-energy of the chiral superfluid, and identify the contribution from the chiral Nieh-Yan anomaly in the low-energy quasirelativistic Weyl superfluid, which leads us to conjecture that the finite temperature NY anomaly term can be similarly universal in general Weyl systems. Finally we compare these results to relativistic Weyl fermions with positive and negative branches at zero momentum.

\section{Torsional anomaly}

For spacetimes with torsion (and curvature), Nieh and Yan \cite{NiehYan1982a,NiehYan1982b,Nieh2007} introduced the 4-dimensional invariant
\begin{equation}
N=\mathcal{T}^a \wedge \mathcal{T}_a - e^a \wedge e^b\wedge R_{ab} \,
\label{N}  
\end{equation}
where $e^a = e^a_{\mu} dx^{\mu}$ is the local tetrad 1-form field and $\mathcal{T}^a = de^a + \omega^a_{\ b} \wedge e^b$ and $R^{ab}=d\omega^a_{b} + \omega^a_{\ c} \wedge \omega^c_{\ b}$, in terms of the tetrad and spin-connection $\omega^a_{\ b} = \omega^a_{\mu b} dx^{\mu}$. As usual, the spacetime metric follows as $g_{\mu\nu} = e^{a}_{\mu} e^{b}_{\nu}\eta_{ab}$, where $\eta_{ab}$ is the local orthonormal (Lorentz) metric. This invariant can be written, using the associated Bianchi identities, as
\begin{equation}
N=dQ \,\,,  \,\, Q=e^a\wedge {\cal T}_a \,.
\label{N2}  
\end{equation}
$N$ is a locally exact 4-from independent from $\textrm{tr} (R \wedge R)$ and the dual of the scalar curvature $\sqrt{g}\mathcal{R}$ in the presence of non-zero torsion. It can be associated with a difference of two topological terms, albeit in terms of an embedding to five dimensions \cite{ChandiaZanelli97a,ChandiaZanelli97b,ChandiaZanelli98}. In terms of four-dimensional chiral fermions on such a spacetime, it has been suggested that this invariant contributes to the axial anomaly, i.e. the anomalous production of the chiral current:
\begin{equation}
\partial_\mu (e j_5^\mu) =  \frac{\Lambda^2}{4\pi^2} N({\bf r},t) \,, \label{eq:NYterm}
\end{equation}
where $e=\det e^a_{\mu}$ and $e j^{\mu}_5$ is the axial current (pseudotensor) density and the non-universal parameter $\Lambda$ has dimension of relativistic momentum (mass) $[\Lambda]=[1/L] = [M]$ and is determined by some ultraviolet (UV) energy scale.

Given the anomaly term \eqref{eq:NYterm}, there has been several attempts to consider the Nieh-Yan anomaly in condensed matter systems with Weyl fermions like \eqref{eq:WeylHamiltonian}, see e.g. \cite{Parrikar2014, SunWan2014, FerreirosEtAl19, Nissinen2019}. However, in non-relativistic systems the relativistic high-energy cut-off $\Lambda$ is not a well defined parameter and, moreover, can be anisotropic. The complete UV theory is, of course, non-Lorentz invariant and the linear, quasirelativistic Weyl regime is valid at much lower scales. Moreover, the anomalous hydrodynamics of superfluid $^3$He at zero temperature suggests that the chiral anomaly is either completely exhausted by the emergent axial gauge field corresponding to the shift of the node or, conversely, the gravitational NY anomaly term arising from the tetrad and spin-connection for local Lorentz invariance along the uniaxial symmetry direction.
It was recently shown in Ref. \onlinecite{Nissinen2019} that the low-energy theory satisfies the symmetries and conservations laws related to an emergent quasirelativistic spacetime with torsion and $\Lambda$ is determined from the UV-scale where the linear Weyl approximation breaks down, as dictated by the underlying $p$-wave BCS Fermi superfluid.  Here we will consider the temperature corrections to the anomaly and the hydrodynamic free-energy in the chiral Weyl superfluid, with the expectation that $\Lambda = T$, $k_B=1$, with some dimensionless prefactor $\gamma$, in units where the tetrads have canonical dimensions of velocity, or the relevant ``speed of light" is set unity. 

In terms of effective low-energy actions, the fully relativistic analogs work unambiguously only for terms in the effective action with dimensionless coefficients. Perhaps the most well-known example being the 2+1-dimensional topological Chern-Simons (CS) terms describing the Quantum Hall effect. Gravitational Chern-Simons terms can be similarly quantized in terms of chiral central charge which has relation to thermal transport and the boundary conformal field theory \cite{Volovik90, ReadGreen01}. The CS action was recently generalized to 3+1d (and higher odd space dimensions) crystalline topological insulators, using so-called dimensionful elasticity tetrads $E$ with dimension $[E]=[1/L]=[M]$. Topological polarization terms can also be written down \cite{SongEtAl2019} by dimensional descent. The ensuing higher dimensional Chern-Simons and polarization terms are expressed as the mixed responses $E\wedge A\wedge d A$ and $E \wedge E \wedge dA$ with quantized dimensionless coefficients \cite{NissinenVolovik2018b, NissinenVolovik2018, SongEtAl2019}. 

Another such example is the temperature correction to curvature effects, with $\delta S_{\rm eff}$ = $\int T^2{\cal R}$ in the low-energy action \cite{VolovikZelnikov2003}. This represents the analog of the gravitational coupling (inverse Newton constant) in the low-energy action, where the curvature scalar ${\cal R}$ is the effective scalar spacetime curvature. Since $[T]^2[{\cal R}]=[M]^4$, the coefficient of this term is dimensionless, and can be given in terms of universal constants: it is fully determined by the number of the  fermionic and bosonic species in the effective theory on flat background, and thus works both in relativistic and non-relativistic systems \cite{VolovikZelnikov2003}. The same universal behavior takes place with the terms describing the chiral magnetic and chiral vortical effects in Weyl superfluid $^3$He-A, where the coefficients are dimensionless \cite{VolovikVilenkin2000,Volovik2003, BasarEtAl14}. Similarly, it has been observed that the coefficient of the $\textrm{tr}(R\wedge R)$ graviational anomaly in chiral Weyl systems affects the thermal transport coefficients in flat space \cite{LandsteinerEtAl11, LoganayagamSurowka12, JensenEtAl13, Landsteiner2014, Lucas2016, GoothEtAl17, StoneKim18}. These coefficients are fundamental, being determined by the underlying degrees of freedom in addition to symmetry, topology and geometry. From this perspective especially, since the NY form is second order in derivatives and can be computed from linear response, our findings are very interesting and warrant further research.

Our goal in the present paper is to separate different $T^2$ contributions in the hydrodynamic free-energy of Weyl superfluids in order to identify the terms responsible for different relativistic phenomena, including a term from the thermal Nieh-Yan anomaly, as well as gravitational terms of the form $\int T^2({\cal R} + \mu^2)$, where $\mu$ is the chiral chemical potential $\mu \ll T$.
 
\section{Temperature correction to the Nieh-Yan term} \label{sec:TempNY}
 
The relativistic zero-temperature anomaly term in the axial current production  $\Lambda^2({\cal T}^a \wedge {\cal T}_a+ e^a\wedge e^b\wedge R_{ab})$ is still not confirmed in general, see however Ref. \onlinecite{Nissinen2019}. On one hand the UV cut-off parameter $\Lambda$ is not well-defined in relativistic field theory with fundamental chiral fermions. On the other hand, such a cut-off is not in general available in non-relativistic matter with quasi-relativistic low-energy chiral fermions and can be anisotropic \cite{Nissinen2019} or even zero. However, the term of the form $\gamma T^2({\cal T}^a\wedge {\cal T}_a+ e^a\wedge e^b\wedge R_{ab})$ has the proper dimensionality $[M]^{4}$, and its prefactor $\gamma$ could be a universal constant in canonical units, being expressed via some invariant related to the degrees of freedom.

For concreteness, we focus on the finite temperature Nieh-Yan anomaly in chiral $p$-wave Weyl superfluid (such as $^3$He-A) with
\begin{equation}
\partial_\mu (ej_5^\mu) = \gamma   T^2 N({\bf r},t) \,,
\label{T2NiehYan}
\end{equation}
and check whether the dimensionless parameter $\gamma$ can be universal. We now use the result obtained by Khaidukov and Zubkov \cite{Zubkov2018} and Imaki and Yamamoto \cite{Imaki2019} for the finite temperature contribution to the chiral current. For a single (complex) chiral fermion, one has for the chiral current
\begin{equation}
j_5^k= -  \frac{T^2}{24}  \epsilon^{kij}\mathcal{T}^0_{ij} \,.
\label{j5}
\end{equation}
We assume that this current can be covariantly generalized  to the 4-current:
\begin{equation}
e j^\mu_5= -  \frac{T^2}{24}  \epsilon^{\mu\nu\alpha\beta} e_{\nu a}\mathcal{T}^a_{\alpha\beta} \,.
\label{j5general}
\end{equation}
Then one obtains the divergence
\begin{equation}
\partial_{\mu} (e j^\mu_5)= -  \frac{T^2}{48}  \epsilon^{\mu\nu\alpha\beta} \mathcal{T}_{a\mu \nu}\mathcal{T}^a_{\alpha\beta}
 \,.
\label{j5nonconservation}
\end{equation}
In the presence of curvature $R(\omega)$, this becomes the temperature correction to the full Nieh-Yan term in Eq. (\ref{T2NiehYan}), where now the non-universal cut-off $\Lambda$ is substituted by the well defined temperature $T$, and the dimensionless parameter $\gamma=1/12$:
\begin{equation}
\partial_\mu (e j^\mu_5) =-  \frac{T^2}{12}   N({\bf r},t) \,.
\label{j5nonconservationR}
\end{equation}
Note it is possible that the local relativistic (Tolman) temperature $T=T_0/\vert e^0_t \vert$ enters the local anomaly, while the constant $T_0$ is the global equilibrium temperature of the condensed matter system \cite{VolovikZelnikov2003}. In \eqref{eq:WeylHamiltonian} we have simply $e^0_t = -1$.

\section{From relativistic physics to chiral Weyl superfluid} \label{sec:superfluid_LL}

In the presence of a finite Weyl node $p_{W}\neq 0$, the chiral anomaly for the chiral current leads to the anomalous production of the linear momentum \cite{Volovik2003, Nissinen2019}.  Even though chiral current is not well-defined at high-energy, the spectral flow of chiral quasiparticles is accompanied by the spectral flow of the linear momentum ${\bf p}_W$ of the Weyl point. In $^3$He-A there are two (spin-degenerate) Weyl points with opposite chirality and opposite momenta, ${\bf p}_{W\pm}=\pm p_F{\hat{\bf l}}$, where $\hat{\bf l}$ is the unit vector of the orbital 
momentum of the superfluid. In particular, the anomalous production of the linear momentum density $n_{\pm}$ at nodes of the two opposite Weyl points sum-up leading to 
\begin{equation}
\dot {\bf P}_{\rm anom} = -p_F \hat{\bf l}\, (\doo_t n_5),
\label{anomalousProductionP}
\end{equation}
where $n_5=n_+-n_-$ is the chiral density. The corresponding quasirelativistic momentum density of the Weyl fermions, valid in the vicinity of the node, is\cite{Nissinen2019}
\begin{equation}
{\bf P}_{\textrm{NY-node}} =- p_F \hat{\bf l} e j^0_5\,.
\label{NYmomentum}
\end{equation}
Thus Eq. \eqref{j5nonconservation} gives the temperature correction to this
anomalous momentum production, leading to a mass current due to thermal Nieh-Yan anomaly. Next we discuss this in detail for the quasiparticles and the superfluid vacuum in the non-relativistic chiral Weyl $p$-wave superfluid, using a Landau level model for the currents \cite{Volovik85,Nissinen2019}. For relativistic Weyl fermions, see \cite{Stone2019}. For considerations of similar temperature effects in chiral Weyl superfluid, see Ref. \cite{KobayashiEtAl18}

\subsection{Nieh-Yan term from the hydrodynamics of chiral Weyl superfluid}

It is known that the hydrodynamics of chiral gapless $^3$He-A experiences momentum anomalies related to the spectral flow through the Weyl points. Let us express Eqs. \eqref{anomalousProductionP}, \eqref{NYmomentum} in terms of the hydrodynamic variables and quasiparticles of the chiral superfluid. The Weyl fermions arise from the BdG Hamiltonian close to the nodes,
\begin{align}
H_{\rm BdG}(-\im \doo) = \left( \begin{matrix} \epsilon(-\im\doo) & \frac{1}{2\im}\{\doo_i,\Delta^i \} \\ \frac{1}{2\im}\{\doo_i,\Delta^{*i} \} & -\epsilon(\im \vek{\doo}) \end{matrix}\right)
\end{align}
Here $\epsilon(\mathbf{p}) = \frac{p^2-p_F^2}{2m^*}$ is the normal state dispersion minus the Fermi level $\mu_F$; $\Delta_i = c_\bot(\hat{\bf m}+i\hat{\bf n})$ is the order parameter in the $p+ip$ chiral superfluid;  
$\hat{\bf l}=\hat{\bf m}\times\hat{\bf n}$ is the unit vector in the direction of the orbital angular momentum of Cooper pairs; anisotropic node velocities are $c_\parallel=v_F$ and $c_\bot =\Delta_0/p_F$, constituting effective speeds of light in Weyl equation along $\hat{\bf l}$
and in transverse directions, respectively. In the weak coupling BCS theory $c_\bot \ll c_\parallel$, in $^3$He-A their ratio is of order $10^{-3}$. In a chiral superconductor we in addition perform the minimal coupling $\epsilon(-\im\doo_i) \to \epsilon(\im D_i) +eA_0$, where $D_i=\doo_i-eA_i$ is the gauge covariant derivative and $A_\mu$ is the electromagnetic potential. Equivalently this amounts to $\mathbf{v_s} \to \mathbf{v}_s-\frac{e\vek{A}}{m}$ in the free-energy.} In what follows we ignore the superfluid velocity (giving rise to spin-connection in addition to tetrads) in the anomaly considerations. Then the only hydrodynamic variable appearing in the torsion is the unit vector of the orbital momentum  $\hat{\bf l}$ \cite{Nissinen2019}. Our results can afterwards then generalized to include the superfluid velocity appearing in the free-energy.

The Weyl nodes are at $\mathbf{p}_{W\pm} = \pm p_F \unitvec{l}$. Expanding the Hamiltonian as $H_{\rm BdG} \simeq \sigma^a e^i_a (\hat{p}-p_W)_i$, the vierbein ${\bf e}^i_a$ in the vicinity of the Weyl point take the form:
\begin{eqnarray}
e^{\mu}_a = \{e^t_{a},{\bf e}^i_a\} &=&
\,\left(\begin{array}{cc}1 & 0\\
 0 & c_\bot \hat{\bf m} \\ 0 & c_\bot \hat{\bf n} \\ 0 &c_\parallel \hat{\bf l}\end{array} \right)  , \quad a = 0,1,2,3.
\end{eqnarray}
For the inverse vierbein ${\bf e}^i_a {\bf e}^a_j = \delta^i_j$ we have
\begin{eqnarray}
e^a_{\mu} = \{e^a_t, {\bf e}^a_i \} &=&
\,\left(\begin{array}{cccc}1 & 0 & 0&0 \\
 0 & \frac{1}{c_\bot}\hat{\bf m}  & \frac{1}{c_\bot}\hat{\bf n} & \frac{1}{c_\parallel}\hat{\bf l}\end{array} \right).  \label{Eu}
\end{eqnarray}
In order to compute the thermal contribution from the quasiparticles in the presence of torsion, we assume that the tetrad gives rise to a constant torsion via $\unitvec{m} = \unitvec{x}$, $\unitvec{n} = \unitvec{y} - T_B x \unitvec{z}$, where $T_B \ll 1$ is a perturbation \cite{Volovik85, Nissinen2019}. In this case, the 3d spectrum organizes into one-dimensional states on 2d Landau levels (LLs) \cite{Volovik85,Volovik2003,Parrikar2014, Stone2019, Laurila20}, where the relevant spatial torsion 
\begin{align}
\frac{1}{2}\epsilon^{ijk}\mathcal{T}^3_{jk} e_3^i = c_{\parallel} \unitvec{l}\cdot ( \nabla \times \frac{\unitvec{l}}{c_{\parallel}}) \equiv T_B,
\end{align}
where $T_B = \unitvec{l}\cdot\nabla\times\unitvec{l} = \mathcal{T}^z_{xy}$, is playing the role of effective magnetic field. Only the gapless lowest LLs are relevant, while the gapped levels, with gap $\sim c_\perp \sqrt{p_FT_B}$  are particle-hole symmetric and cancel out \cite{Volovik85, BalatskiiEtAl86}. We approximate $\epsilon(\mathbf{p}) = \frac{p^2-p_F^2}{2m} \simeq \frac{p_z^2-p_F^2}{2m} = \epsilon(p_z)$, strictly valid when $p_{\bot} \ll m c_{\perp}$ which coincides with the linear Weyl regime $\epsilon_{p_z} = \pm v_F(p_z-p_F)$. However, we can fix the anisotropic dispersion to $\epsilon(p_z)$ for all momenta to obtain a convenient UV-complete model, for which the following analysis for the total vacuum current is valid, as long as the correct cutoff for the linear Weyl regime is maintained for $^3$He-A, see Appenix \ref{Appendix2}. The dispersion of the lowest LL becomes $E_{n=0} = -\textrm{sgn}(p_z T_B)\epsilon(p_z)$, while density of states per lowest LL per three-momentum $p_z$ becomes $n_{\rm LL}(p_z) = \frac{\abs{p_z T_B}}{4\pi^2}$. The lowest LL is particle-like (hole-like) for $p_W = \mp p_F\unitvec{l}$. For more on this non-relativistic model with torsion, see \cite{Volovik85, BalatskiiEtAl86,Nissinen2019, Laurila20}

The total (chiral) quasiparticle current along $\unitvec{l}=\frac{e^i_3}{v_F}\approx \unitvec{z}$ becomes
\begin{align}
\mathbf{P}^{\rm qp}\cdot\unitvec{l} &= -2 \int_{0}^{\infty} N_{\rm LL}(p_z) dp_z~  p_z n_{F}(\epsilon_{p_z}-\mu_F) \nonumber\\
&=-\left(\frac{p_F^3}{6\pi^2}+\frac{p_F T^2}{6 c_{\bot}^2} \right) \unitvec{l}\cdot\nabla\times\unitvec{l} \label{eq:LL_current} \\
&= \vek{j}^{\rm vac}_{\rm anom\parallel} + \vek{j}^{\rm qp}_{\rm anom\parallel}(T) \nonumber. 
\end{align}
where $n_{F}(x) =(e^{\beta x} +1)^{-1}$ is the quasiparticle distribution function and a factor of two comes from the spin-degeneracy. In order to compute the integral, 
we have used $T\ll\Delta_0 \ll \mu_F$ as well as the linear expansion $\epsilon(\Delta p_z) \approx v_F(\Delta p_z c_{\perp}/v_F) \approx T$ with anisotropic momentum scaling, where close to the nodes $\Delta p_z \sim \frac{T}{\Delta_0}p_F = \frac{T}{c_{\perp}}$ for both $T, \epsilon \ll mc_\perp^2$, the cutoff of the linear Weyl regime in $^3$He-A. See Appendix \ref{Appendix2} and \cite{Nissinen2019,Laurila20} for more details.

From the above we conclude that $\vek{j}^{\rm vac}_{\rm anom}$ is the anomalous superfluid current from filled quasiparticle states, see Eq. \eqref{eq:j-current} below,
\begin{align}
\vek{j}^{\rm vac}_{\rm anom\parallel} = -\frac{p_F^3}{6\pi^2}\unitvec{l}(\unitvec{l}\cdot \nabla \times \unitvec{l}) = -\frac{C_0}{2}\unitvec{l}(\unitvec{l}\cdot \nabla \times \unitvec{l}), \label{eq:vacuum_j}
\end{align}
whereas the finite temperature contribution to the quasiparticle momentum is
\begin{align}
\vek{j}^{\rm qp}_{\rm anom\parallel}(T) = -\frac{p_F T^2}{6 c_{\perp}^2}\unitvec{l}(\unitvec{l}\cdot \nabla \times \unitvec{l}) \label{eq:Tcorrection}.
\end{align}
and arises due to thermal normal component close to the nodes. These contributions to the anomalous vacuum current from the perspective of the superfluid are analyzed in the next subsection.
 
The relativistic anomaly results in Sec. \ref{sec:TempNY} and the Landau level argument for the chiral superfluid suggest in addition the following temperature correction to the anomalous momentum in $^3$He-A, in the vicinity of the node, from the thermal Nieh-Yan anomaly \cite{Nissinen2019}:
\begin{align}
{\bf P}_{\textrm{NY-node}}(T) =-p_F \hat{\bf l}\, n_5(T)= \frac{p_F T^2 }{12  c_\bot^2}\hat{\bf l}(\hat{\bf l}\cdot(\nabla\times \hat{\bf l})).
\label{deltaPanomalous}
\end{align}
With the Landau level approximation, the quasiparticle density $n_5(T)$ between the two Weyl nodes is computed from a similar integral as Eq. \eqref{eq:LL_current},
\begin{align} 
n_5(T) &= -2 \int_{0}^{\infty} N_{\rm LL}(p_z) dp_z \bigg[n_F(\epsilon_{p_z}-\mu_F) -\Theta(\mu-\epsilon_{p_z})\bigg] \nonumber\\
&= -\frac{T_B}{2\pi^2}  \int_0^{\infty} dx ~x n_{F}(x) \nonumber\\
&= - \frac{T^2}{12c_{\perp}^2} (\unitvec{l}\cdot\nabla\times\unitvec{l}). \label{eq:LL_density}
\end{align}
Similar to the total current, this is the contribution from the linear spectrum close to the nodes with $\Delta p_{\parallel}\sim(\frac{c_{\perp}}{c_{\parallel}})\frac{T}{c_{\bot}}$ due to thermal fluctuations, which renormalizes the velocity coefficient and requires the cutoff of the Weyl regime relevant for the full dispersion, such as $^3$He-A. The thermal fluctuations for the superfluid are discussed in Appendix \ref{Appendix1}.

What have been calculated above are actually the temperature dependence of the anomalous torsional conductivities of the superfluid, i.e. then chiral momentum-density and number densities in response to $T_B$, see e.g. \cite{Landsteiner2014}. Note in particular that the leading torsional contribution from the NY anomaly in Eqs. \eqref{eq:LL_current}, \eqref{eq:LL_density} from thermal fluctuations is suppressed by the the factor $c_{\bot}^{-2}$, in contrast to the antisymmetric torsion $S = \frac{1}{2}\epsilon^{ijk} e^3_i \mathcal{T}^3_{jk} = \frac{T_B}{c_\parallel^2}$ from \eqref{Eu}, as was found in Ref. \cite{Nissinen2019} without the Landau level approximation to the anomaly. It is also intriguing to see that the non-relativistic thermal integrals in Eqs. \eqref{eq:LL_current},\eqref{eq:LL_density} with explicit non-relativistic vacuum regulatization due to the filled quasiparticle states coincide with similar expression for relativistic Weyl fermions with positive and negative branches \cite{LoganayagamSurowka12, JensenEtAl13, BasarEtAl14, Landsteiner2014, Zubkov2018, Imaki2019}, see Appendix \ref{Appendix3}.

\subsection{Anomalous vacuum current}

The hydrodynamic anomaly in momentum conservation arises between the quasiparticles and vacuum mass current $\vek{P} = \vek{j}$ of the superfluid, which at $T=0$ has the following general form \cite{Cross1975, VollhardtWolfle, Volovik2003}:
\begin{align}
\mathbf{j} = \rho \mathbf{v}_s + \frac{1}{4m} \nabla \times (\rho \unitvec{l}) - \frac{C_0}{2} \unitvec{l} (\unitvec{l} \cdot (\nabla \times \unitvec{l}) )\,, 
\label{eq:j-current}
\end{align}
where the last term is anomalous with the parameter $C_0$ from the combined orbital-gauge symmetry of the superfluid \cite{VolovikMineev81}, that fully determines the axial anomaly in the system due to the Weyl quasiparticles of the superfluid \cite{Volovik2003}
\begin{align}
C_0(T=0) = p_F^3/3\pi^2 = \rho,
\end{align}
where we ignore corrections of the order of $(\Delta_0^2/E_F^2) = (c_{\perp}/c_{\parallel})^2 \ll 1$ at zero temperature to the density from pairing ($\rho-C_0 \sim \rho (\Delta_0^2/E_F^2)$). Notably, this term exists only on the weak coupling side of the topological BEC-BCS Lifshitz transition, where the pair of the Weyl points with ${\bf p}_{W\pm}=\pm p_F{\hat{\bf l}}$ appears in the quasiparticle spectrum \cite{VolovikMineev81, Volovik2003}. 

Here we are interested in the temperature correction to the anomaly, which may come from the Nieh-Yan term.
The extension of the anomalous current  to nonzero temperature gives 
\begin{equation}
 {\bf P}_{\rm anom}(T) = -\frac{1}{2}C_0(T)  \hat{\bf l}(\hat{\bf l}\cdot(\nabla\times \hat{\bf l})
 \,,
\label{Cross}
\end{equation}
where according to Cross \cite{Cross1975,VollhardtWolfle}, the anomalous parameter $C_0(T)$
has the following temperature dependence at low $T\ll T_c$:
\begin{equation}
C_0(T)  =C_0(0) - T^2   \frac{p_F}{6 c_\bot^2} \left(1+ \frac{m^*}{m} \right)\,. 
\label{C}  
\end{equation}
Here  $m^*$ is the effective mass of quasiparticles in the normal Fermi liquid, which differs from the bare mass $m$ of the $^3$He atom due to the Fermi liquid corrections. In Eqs. \eqref{eq:vacuum_j} and \eqref{eq:Tcorrection} we have neglected the Fermi-liquid corrections due to interactions:  An additional factor $\frac{1}{2}(\frac{m^*}{m}-1)(\rho_n^{(0)}/\rho)$ arises from the reduced quasiparticle momentum flow, due to Galilean invariance. The current becomes \cite{Cross1975}
\begin{align}
\vek{j}_{\rm anom} = \frac{1}{1+\frac{1}{3} F^s_1(\rho^{(0)}_{n\parallel}/\rho)} \frac{m^*}{m} \vek{j}^{(0)}_{\rm anom} \label{eq:backflow}
\end{align}
where the Landau parameter $\frac{1}{3}F_1^s = \frac{m^*}{m}-1$ and $\rho_{n\parallel, \perp}^{(0)}(T)$ are the bare thermal quasiparticle densities without Fermi-liquid corrections along and perpendicular to $\unitvec{l}$. They are given by, see the Appendix \ref{Appendix1},
\begin{align}
\rho^{(0)}_{n\parallel} = \frac{\pi^2 T^2}{\Delta_0^2}\rho, \quad \rho^{(0)}_{n\bot} = \frac{7\pi^4T^4}{15 \Delta_0^4}\rho, \label{eq:rhoT}
\end{align}
where by Galilean invariance, $\rho\delta_{ij} = \rho_{s ij} + \rho_{n ij}$ at all temperatures. From Eqs. \eqref{eq:backflow}, \eqref{eq:rhoT} we gather
\begin{align}
\frac{C_0(T)}{2} = \frac{p_F^3}{6\pi^2} - \frac{p_F T^2}{6c_{\bot}^2} -\frac{p_F T^2}{12 c_{\perp}^2}\left(\frac{m^*}{m}-1\right) + O(T^4), \label{eq:NY_Fermi_corrections}
\end{align}
which is the Fermi-liquid corrected result Eq. \eqref{eq:backflow} and corresponds to the result \eqref{eq:Tcorrection} from the reduction of superfluid density when the Fermi-liquid corrections are ignored.

We summarize these findings as follows. While Eq. \eqref{Cross} with temperature corrections Eq. \eqref{C} looks similar to Eq. \eqref{deltaPanomalous}, $\vek{P}_{\rm anom}(T)$ represents the consistent anomaly vacuum momentum density from all filled states which depends on the non-fundamental parameters $m^*$ and $m$ via the Fermi-liquid corrections, while we expect that the $T^2$-contribution to the
(covariant) NY anomaly arise from contributions close to the linear Weyl nodes and should contain fundamental and universal prefactors. 

Concerning the vacuum momentum current of the superfluid, one reason is that there are several different $T^2$ contributions to the current and free energy in the chiral superfluid, and they correspond to phenomemena with different origins and scales.  Although of similar form in terms of the low-energy Goldstone variables of the superfluid, they can expressed in relativistic form  with fundamental parameters only when carefully keeping track of each individual contribution to avoid double counting. In particular, in the next section we shall see that  the non-fundamental parameters $m^*$ and $m$ do not enter the free energy or the final results, when experessed in terms of the correct relativistic variables valid in the quasirelativistic low-energy theory. 

\section{Relativistic corrections to free-energy}

Let us consider the $T^2$-corrections in the free energy which are second order in derivatives containing combinations of $(\hat{\bf l}\cdot {\bf v}_s)$ and $(\hat{\bf l}\cdot (\nabla\times\hat{\bf l}))$, neglecting all higher order $O(T^4, \doo^3)$ terms.  These terms can be distributed into three groups, which have different dependence on $m^*$ and $m$:
\begin{align}
F&=F_1+F_2+F_3 \,,
\label{123}  
\\
F_1[\unitvec{l},\vek{v}_s]&=\frac{p_F}{12} \frac{T^2}{c_\perp^2}(\hat{\bf l}\cdot {\bf v}_s)(\hat{\bf l}\cdot (\nabla\times\hat{\bf l})),
\label{1}  
\\
F_2[\unitvec{l},\vek{v}_s]&=- \frac{p_Fm^*}{96m^2} \frac{T^2}{c_\perp^2}
\left(4m(\hat{\bf l}\cdot {\bf v}_s)-(\hat{\bf l}\cdot (\nabla\times\hat{\bf l}))\right)^2,
\label{2}  
\\
F_3[\unitvec{l}]&=- \frac{p_F}{288m^*} \frac{T^2}{c_\perp^2}(\hat{\bf l}\cdot (\nabla\times\hat{\bf l}))^2,
\label{3}  
\end{align}
The relativistic form of each of these free-energy contributions arises separately as follows. 

The term $F_3$ in Eq. \eqref{3} describes the universal temperature correction to the Newton gravitational coupling, which depends on the number of fermionic species \cite{VolovikZelnikov2003}:
\begin{equation}
F_3[\mathcal{R},T]=- \frac{v_F}{288} \frac{T^2}{c_\perp^2}(\hat{\bf l}\cdot (\nabla\times\hat{\bf l}))^2
=\frac{T^2}{144}\sqrt{-g} \cal{R}\,.
\label{NewtonG}  
\end{equation}
Note that being exressed in terms of the scalar curvature and metric determinant $\sqrt{-g} =1/v_Fc_\perp^2$, this term becomes universal: it does not contain the microscopic parameters of the system: $p_F$, $m$ and $m^*$. The prefactor is fully determined by the number of fermionic species.

The term $F_2$ in Eq. \eqref{2} is expressed in terms of the combination $\unitvec{l}\cdot  \vek{v} = \hat{\bf l}\cdot {\bf v}_s-\frac{1}{4m}\hat{\bf l}\cdot (\nabla\times\hat{\bf l})$, proportional to the ground state current which does not receive corrections due to Galilean invariance. Here the velocity ${\bf v}={\bf j}/\rho$ is the velocity of the ``total quantum vacuum", where ${\bf j}\equiv {\bf j}(T=0)$ is the total vacuum current in Eq. \eqref{eq:j-current} at $T=0$.
We conclude that the $F_2$ contribution gives the temperature $T$ and chemical potential $\mu$ correction to the free energy of the gas of chiral fermionic particles in the limit $|\mu| \ll T$ (see Eqs. (9.12) and (10.42) in Ref. \cite{Volovik2003}):
\begin{eqnarray}
F_2[\mu,T] =- \frac{p_Fm^*}{96m^2} \frac{T^2}{c_\perp^2}
\left(4m\,\hat{\bf l}\cdot {\bf v}_s-\hat{\bf l}\cdot (\nabla\times\hat{\bf l})\right)^2
=\nonumber
\\
=-\frac{T^2}{6} \sqrt{-g} \mu^2  \,,
\label{A0}  
\end{eqnarray}
where the chiral chemical potential  of the superfluid is determined by the Doppler shift
\begin{equation}
\mu_R=-\mu_L =p_F {\bf v}\cdot \hat{\bf l}= p_F \hat{\bf l}\cdot {\bf v}_s-(p_F/4m)\hat{\bf l}\cdot (\nabla\times\hat{\bf l})
 \,.
\label{A0mu}  
\end{equation}
The final version of Eq. \eqref{A0} does not contain microscopic parameters. The Eq. \eqref{A0} also gives rise to the mass of ``photon" in $^3$He-A \cite{Volovik1998}. In principle, the chiral chemical potential may also serve as the $\Lambda$ parameter in the Nieh-Yan term, see however Appendix \ref{Appendix3}.

Finally $F_1$ in Eq.(\ref{1})  is the term in free energy, which gives rise to the vacuum current without any factors of $\frac{m^*}{m}$ with contribution from the Nieh-Yan thermal anomaly in Eqs. \eqref{deltaPanomalous}, \eqref{C}, \eqref{eq:NY_Fermi_corrections}:
\begin{align}
{\bf P}_{\rm anom}(T) &= \frac{\delta F_1}{\delta {\bf v}_s}=\frac{p_F}{12} \frac{T^2}{c_\perp^2}\hat{\bf l}(\hat{\bf l}\cdot (\nabla\times\hat{\bf l})) \label{J}
\\
&\simeq -p_F n_{5}(T)\hat{\bf l}, \nonumber
\end{align}
where the last equality was derived in Eq. \eqref{eq:LL_density}. We stress however that $\vek{P}_{\rm anom}(T)$ represents the total anomalous superfluid current, whereas the right-hand side is equal to the quasiparticle momentum density due to thermal fluctuations close to the node.

\section{Conclusions} 

We discussed the possibility of thermal Nieh-Yan anomaly where the role of the non-universal dimensionful UV cut-off $\Lambda$ is played by the temperature IR scale, and the dimensionless prefactor $\gamma$ in the anomaly is universal.  We identified a contribution from this anomaly in the known low-temperature corrections of non-relativistic chiral $p$-wave Weyl superfluid (or superconductor). In this system, the anomaly results from thermal effects of the linear Weyl spectrum at finite momentum in the presence of an explicit vacuum of filled quasiparticles, although the end results are similar to relativistic fermions when interpreted carefully in terms of the anisotropy and cutoff of the quasirelativistic Weyl regime.

What we calculated, via the anisotropic Landau level model with non-relativistic symmetries in Sec. \ref{sec:superfluid_LL}, are actually the temperature dependence of the anomalous torsional conductivities $\sigma_{\mathcal{T}^a}$ of the quasiparticle system, i.e. the chiral momentum-density and number densities (at the nodes at finite momenta $\pm p_F$) in response to the spacelike torsion $T_B= \frac{1}{2}\epsilon^{ijk}e^3_i \mathcal{T}^3_{jk}$. Namely, for example, 
\begin{align}
\langle P^a \rangle =\langle e^a_i T^{0i} \rangle = e^a_i \frac{\sigma_{\mathcal{T}^b}(T)}{2} \epsilon^{0ijk} T^b_{jk} = \delta^{a3}\sigma_{T_B} T_B,
\end{align}
$P^i = T^{0i}$, where $T^{\mu\nu}$ is the stress-tensor, which are of second order in derivates and can be calculated in linear response, e.g. Kubo formulas, to the background tetrads \cite{LandsteinerEtAl11, Landsteiner2014, BradlynRead15, Gromov15}. A similar momentum anomaly in anisotropic system with non-relativistic symmetries was also considered in \cite{Copetti20}. For the zero temperature case, see \cite{Nissinen2019,Laurila20} as well as \cite{Stone2019, Stone2019b} for a relativistic model related to Weyl semi-metals. We, however, stress that the universal gravitational NY anomaly and thermal physics we discussed arise in flat space from the geometric background fields in the low-energy quasiparticle Hamiltonian. These tetrads arise universally in all Weyl systems \eqref{eq:WeylHamiltonian} and couple to the momentum \cite{Parrikar2014, ShapourianEtAl15} as in gravity. The connection of our results and the relation of the gravitational NY anomaly, with the coefficient $\gamma$, to thermal transport in Weyl system should be further elucidated \cite{Luttinger64, LandsteinerEtAl11, LoganayagamSurowka12, JensenEtAl13, Lucas2016, GoothEtAl17, KobayashiEtAl18}.

Detailed consideration of the temperature dependent anomaly terms in the hydrodynamics of the non-relativistic $p$-wave chiral superfluid with quasirelativistic Weyl fermions demonstrates that in the hydrodynamics of this liquid there are several $T^2$ terms, which can be assigned to different emergent relativistic phenomena, both anomalous and non-anomalous. In particular, we identified and discussed the term in the vacuum momentum corresponding to the (consistent) thermal Nieh-Yan anomaly. As expected this term originates from thermal fluctuations close to the linear nodes with emergent quasirelativistic torsion with anisotropy. Note that in terms of the  superfluid and Weyl fermions, the $T=0$ anomalous  vacuum contribution to the current can be assigned only to a non-local action \cite{Volovik1986c}.  We showed how the various $T^2$ low-temperature corrections can be written aslow-energy relativistic terms with dimensionless prefactors, which do not seem to depend on microscopic physics, but are fully determined by geometry, topology and the number of fermionic and bosonic quantum fields. Detailed comparison of the finite temperature superfluid hydrodynamics with Fermi-liquid corrections to the anomalous quasiparticle axial current production in the presence of arbitrary textures and superfluid velocity remains to be identified \cite{Volovik85, Combescot86} . See however Ref. \onlinecite{Nissinen2019} for the zero temperature case.

\emph{Note added:} After the initial submission of this manuscript as a preprint, arXiv:1909.08936v1, with the predicted $T^2$-contribution to the NY anomaly, Eqs. \eqref{j5nonconservationR}, \eqref{J}, the recent preprints \cite{Stone2019, Stone2019b, Ojanen2019} discussing related torsional anomaly phenomena at finite temperatures appeared. General aspects of the temperature anomaly in Weyl materials were further discussed in the short paper \cite{NissinenVolovik2019}. In particular, the result Eq. \eqref{j5nonconservationR} has been confirmed in Ref. \onlinecite{Stone2019} by a direct calculation of the spectral flow of relativistic Landau levels in the presence of a constant torsional magnetic field $T^3_{\mu\nu}$ \cite{Volovik85, BalatskiiEtAl86, Parrikar2014} at finite temperature. Here similar computations for the non-relativistic Weyl superfluid in Eqs. \eqref{eq:LL_current} and \eqref{eq:LL_density} give corresponding results. While the current manuscript was being finalized, also the paper \cite{Imaki20} appeared discussing the anomaly for relativistic fermion at finite temperature and chemical potential. 

\emph{Acknowledgements:} GEV thanks Mike Zubkov for discussions. JN thanks Z.-M. Huang for correspondence and T. Ojanen for discussions.
This work has been supported by the European Research Council (ERC) under the European Union's Horizon 2020 research and innovation programme (Grant Agreement No. 694248).

\appendix

\section{Vacuum regularization and thermal integrals}\label{appendix}
In this appendix we shortly review the low-temperature corrections to the chiral superfluid and the quasirelativistic thermal integrals utilized in the main text. Although the temperature and Fermi-liquid corrections in Sec. \ref{Appendix1} chiral superfluid are well-known \cite{Cross1975, VollhardtWolfle, Volovik2003}, it is of interest to compare them to the chiral quasirelativistic Landau level model with torsion and explicit UV completion \cite{Volovik85, Nissinen2019, Laurila20} Sec. \ref{Appendix2}, as well to relativistic Weyl fermions with positive and negative branches \cite{LoganayagamSurowka12, BasarEtAl14, Landsteiner2014} Sec. \ref{Appendix3}. The last system has particles and antiparticles at finite chemical potential \emph{counted from the node} and finite temperature with vacuum contributions subtracted, in contrast to non-relativistic systems at finite chemical potential. 

\subsection{Temperature corrections to normal and superfluid density in chiral superfluid}\label{Appendix1}
We start with the corrections to the superfluid and normal density. The anisotropic chiral $p$-wave superfluid density is 
\begin{align}
\rho^{(0)}_{n ij} = 3\rho \int d\Omega~ \unitvec{k}_i \unitvec{k}_j \int_{-\infty}^{\infty} d\epsilon~ \left(-\frac{\doo n_F(E_{\vek{k}})}{\doo E_\vek{k}}\right)_{k=k_F}
\end{align}
where $n_F(x) = (e^{\beta x}+1)^{-1}$ with $\beta = 1/T$ and $E_{\vek{k}} = \sqrt{\epsilon_{\vek{k}}^2 + \abs{\Delta_{\vek{k}}}^2}$ is the quasiparticle energy with normal state dispersion $\epsilon_{\vek{k}} = \frac{k^2}{2m}-\mu_F$. We compute the anisotropic contributions from, with $\omega_{n} =\pi T(2n+1)$,
\begin{align}
\int_{-\infty}^{\infty} d\epsilon~ &\left(-\frac{\doo n_F}{\doo E_{\vek{k}}}\right)_{k_F} =  \int_{-\infty}^{\infty} d\epsilon~ T\sum_{n=-\infty}^{\infty} \frac{\omega_n^2-E_{k_F}^2}{(\omega_n^2+E_{k_F})^2} \nonumber\\
& =  \pi T \sum_{n=-\infty}^{\infty} \frac{-\abs{\Delta_{k_F}}^2}{(\omega_n^2+\abs{\Delta_{k_F}}^2)^{3/2}} \label{eq:FermiDerivative}
\end{align}
The dimensionless summation variable is $x_n = \frac{\pi T}{\Delta_0}(2n+1)$, giving for $T\ll \Delta_0$
\begin{align}
\frac{\rho^{(0)}_{n\parallel}}{3 \rho} &=  \frac{\pi T}{\Delta_0} \sum_{n=-\infty}^{\infty}\int_{-1}^{1} du \frac{-u^2(1-u^2)}{(x_n^2+(1-u^2))^{3/2}} \nonumber \\
&=\frac{\pi T}{2\Delta_0} \sum^{\infty}_{n=-\infty} \left[3 \abs{x_n}-(3x^2_n+1)  \arctan\left(\frac{1}{\abs{x_n}}\right) \right] \nonumber\\
&=\frac{\pi T}{\Delta_0} \sum_{n=-\infty}^{\infty} \left[4 \abs{x_n}-(3 x^2_n-1)\frac{\pi}{2} + O(x_n^4)\right] 
\end{align}
and
\begin{align}
\frac{\rho^{(0)}_{n\perp}}{3\rho}&=  \frac{\pi T}{\Delta_0} \sum_{n=-\infty}^{\infty}\int_{-1}^{1} \frac{du}{4} \frac{-(1-u^2)^2}{(x_n^2+(1-u^2))^{3/2}} \nonumber \\
&=\frac{\pi T}{4\Delta_0} \sum^{\infty}_{n=-\infty} \left[3 \abs{x_n}-\frac{2\abs{x_n}}{1+x_n^2} \right. \nonumber\\
&\hspace{3cm} \left.+(1-3x^2_n)  \arctan\left(\frac{1}{\abs{x_n}}\right) \right] \nonumber\\
&=\frac{\pi T}{4\Delta_0} \sum_{n=-\infty}^{\infty} \left[-\frac{16}{3} \abs{x_n}^3+(3 x^2_n-1)\frac{\pi}{2} + O(x_n^5)\right] \nonumber
\end{align}
We use the regularizations, $\zeta_a(s) = \sum_{n=0}^{\infty} (n+a)^s$,
\begin{align}
\sum_{n=-\infty}^{\infty} &= 1 + 2 \sum_{n=1}^{\infty} = 1+2\zeta(0) = 0,\nonumber\\
\sum_{n=-\infty}^{\infty} \abs{x_n} &= \frac{4\pi T}{\Delta_0}\zeta_{1/2}(-1) = \frac{\pi T}{6\Delta_0},\nonumber\\ 
\sum_{n=-\infty}^{\infty} x^2_n &= \frac{8 \pi^2T^2}{\Delta_0^2}\zeta_{1/2}(-2) = 0,\\
\sum_{n=-\infty}^{\infty} \abs{x_n}^3 &= \frac{16 \pi^3T^3}{\Delta_0^3}\zeta_{1/2}(-3) = -\frac{7\pi^3T^3}{60\Delta_0^3} \nonumber
\end{align}
These give the result in Eq. \eqref{eq:rhoT}. Similarly $\rho_{sij}(T) = \rho\delta_{ij} -\rho_{nij}(T)$ by Galilean invariance, valid at all temperatures with or without Fermi liquid corrections.

From the superfluid and normal densities, the conclusion is that only anisotropic momenta of the order of $\Delta p_{\perp} \sim \frac{T}{\Delta_0} p_F = \frac{T}{c_{\perp}}$ and $\Delta p_{\parallel} = (\frac{c_\perp}{c_{\parallel}}) \frac{T}{c_{\perp}}$, $c_{\parallel} = v_F$ contribute in the vicinity of the gap nodes to the thermal fluctuations $\epsilon \sim T$. 

\subsection{Quasirelativistic chiral fermions with torsion} \label{Appendix2}

The quasirelativistic model for the chiral superfluid with anisotropic dispersion is, with $\unitvec{l}\simeq \unitvec{z}$,
\begin{align}
\epsilon_{p} = \frac{p^2}{2m}-\mu_F \to \epsilon_{p_z} \equiv \frac{p_z^2}{2m}-\mu_F \label{Aeq:dispersion}
\end{align}
which is valid for real $^3$He-A for $E \ll mc_{\perp}^2$, $c_{\perp} \equiv \Delta_0/p_F$. This limit coincides with linear Weyl regime, where $\epsilon_p = \pm c_{\parallel}(p_z-p_F)$, $c_{\parallel}\equiv v_F$, but the model \eqref{Aeq:dispersion} allows calculations of Landau levels and the filled quasiparticle states up till $p_z=0$ that contribute to the superfluid vacuum. In case of the anomalous current, only the momentum along $\unitvec{l} \simeq \unitvec{z}$ contributes justifying the approximation as long as the cutoff $T\ll \Delta_0$ and $E \ll mc_\perp^2 =\frac{c_{\perp}}{c_\parallel} \Delta_0$ are maintained. In particular, this requires that
\begin{align}
\epsilon_p(p_F+\Delta p) \to \epsilon_{p_z}(p_F+\Delta p_{\parallel}) \sim c_\parallel \left(\frac{c_{\perp}}{c_{\parallel}}\right) \frac{T}{c_\perp} \sim T \label{Aeq:scaling}
\end{align} 
by the equipartition principle. The momenta (and energies) should be anisotropically scaled in the simplified model, or equivalently $\epsilon_{p_z} \to (\frac{c_\perp}{c_\parallel})\epsilon_{p_z}$ for $^3$He-A with torsion. The density of state per $p_z$ per Landau level is
\begin{align}
N_{\rm LL}(p_z) = \frac{\abs{p_zT_{B}}}{4\pi^2}
\end{align}
where $T_{B} = \unitvec{l}\cdot \nabla\times\unitvec{l}$. For more on this anisotropic chiral fermion with torsion, see Ref. \onlinecite{Nissinen2019} and the forthcoming paper \cite{Laurila20}.

In case of the Landau level calculation, the quasirelativistic integrals are given by the lowest Landau level (LLL) contribution, which is particle- and hole-like for $p_z<0$ and $p_z>0$, respectively. In the chiral superfluid, the quasiparticles are spin-degerate Majorana-Weyl excitations. It follows that the quasiparticle momentum along is the $\unitvec{l}=\unitvec{z}$ direction is
\begin{align}
\mathbf{P}_{\rm qp}\cdot\unitvec{l} &= -2 \int_{0}^{\infty} N_{\rm LL}(p_z) dp_z p_z n_{F}(\epsilon_{p_z}-\mu_F)\nonumber\\
&= -\frac{2 T_B}{4\pi^2}  \int_0^{\infty} dp_z p_z^2 n_{F}(\epsilon_{p_z}-\mu_F) \nonumber\\
&= -\frac{T_B}{2\pi^2} \bigg[ \int^{p_F}_0 dp_z p_z^2 +\frac{4p_F}{c_{\bot}^2}\int_0^{\infty}dx~ x n_F(\frac{x}{T}) \bigg] \nonumber\\
&=-\left(\frac{p_F^3}{6\pi^2}+\frac{p_F T^2}{6 c_{\bot}^2} \right) \unitvec{l}\cdot\nabla\times\unitvec{l} \nonumber \\
&= \vek{j}^{\rm vac}_{\rm anom\parallel} + \vek{j}^{\rm qp}_{\rm anom\parallel}(T). 
\end{align}
where we have used $T\ll \Delta \ll \mu_F$ and the linearization of the spectrum close to $\mu_F$ with the anisotropic scaling \eqref{Aeq:scaling} which applies for (weak-coupling) $^3$He-A. The vacuum contribution is
\begin{align}
\vek{j}^{\rm vac}_{\rm anom\parallel}= -\frac{T_B}{2\pi^2} \int_{0}^{\infty} dp_z p_z^2\Theta(p_F-p_z)
\end{align}
whereas the quasiparticle current due to the thermal fluctuations is
\begin{align}
\vek{j}^{\rm qp}_{\rm anom\parallel}(T) = -\frac{p_F T^2}{6 c_{\perp}^2}\unitvec{l}(\unitvec{l}\cdot \nabla \times \unitvec{l}).
\end{align}

Similarly, the thermal density of states is the vicinity of the nodes for $T\ll \Delta_0$ is
\begin{align}
n_5(T) &= -2 \int_{0}^{\infty} N_{\rm LL}(p_z) dp_z \bigg[n_F(\epsilon_{p_z}-\mu_F) -\Theta(\mu-\epsilon_{p_z})\bigg] \nonumber\\
&= -\frac{T_B}{2\pi^2}  \int_0^{\infty} dx ~x n_{F}(x) \nonumber\\
&= - \frac{T^2}{12c_{\perp}^2} (\unitvec{l}\cdot\nabla\times\unitvec{l}). \label{Aeq:LL_density}
\end{align}
where the $T=0$ vacuum contribution is substracted, since only contribution from the vicinity of the nodes to the chiral density is well-defined. The relevant thermal integral is similar to Eq. \eqref{eq:FermiDerivative}
\begin{align}
\int_0^{\infty} dx~ x n_F(x) = \frac{1}{4} \int^{\infty}_{-\infty} x^2\left(-\frac{\doo n_F(x)}{\doo x}\right) \\
=  4\pi^2 T^2 \zeta_{1/2}(-1) = \frac{\pi^2 T^2}{6}
\end{align}

\subsection{Relativistic chiral fermions}\label{Appendix3}
Now we compare these results with relativistic Weyl fermions with both positive and negative energy branches, i.e. particles and antiparticles with the node at $p_z=0$. At chemical potential $\mu$ at the node, the corresponding momentum along $\unitvec{l}$ is
\begin{align}
\mathbf{P}_{\rm W}\cdot\unitvec{l} &= \int_{-\infty}^{\infty} N_{\rm LL}(p_z) dp_z p_z \bigg[n_{F}(\epsilon_{p_z}-\mu)-\Theta(-p_z)\bigg] \nonumber\\
&= -\frac{T_B}{4\pi^2}  \int_0^{\infty} dp_z p_z^2 \bigg[n_{F}(\epsilon_{p_z}-\mu)- n_F(\epsilon_{p_z}+\mu)\bigg] \nonumber\\
&= -\frac{T_B}{4\pi^2} \bigg[ \int^{\mu}_0 dp_z p_z^2 +4\mu\int_0^{\infty}dx~ x n_F(\frac{x}{T}) \bigg] \nonumber\\
&=-\left(\frac{\mu^3}{12\pi^2}+\frac{\mu T^2}{12} \right) \unitvec{l}\cdot\nabla\times\unitvec{l},
\end{align}
per \emph{single} node, in units where $\epsilon_{p_z} = p_z$. The thermal chiral density, corresponding to the left-handed node,
\begin{align}
n_{W-}(T) &= -\int_{-\infty}^{\infty} N_{\rm LL}(p_z) dp_z \bigg[n_{F}(\epsilon_{p_z}-\mu)-\Theta(-p_z)\bigg] \nonumber\\
&= -\frac{T_B}{4\pi^2} \int_{0}^{\infty} dp_z p_z\bigg[n_F(\epsilon_{p_z}-\mu)+n_{F}(\epsilon_{p_z}+\mu)\bigg] \nonumber \\
&= (-\frac{\mu^2}{8\pi^2}-\frac{T^2}{24})\unitvec{l}\cdot\nabla\times\unitvec{l}.
\end{align}
In order to compute the integral, we have used $n_F(x)=1-n_F(-x)$ as well as the linear expansion close to the nodes for $T\ll \Delta_0$. Intriguinly, the structure of this momentum-density and density contribution are exactly of the same form \cite{LoganayagamSurowka12, BasarEtAl14, Landsteiner2014, Zubkov2018, Imaki2019}. The difference is, of course, that $\mu$ is counted from the node and has different sign for positive energy solutions corresponding to the antiparticles and that the contribution of filled vacuum states is substracted. Lastly, we note the dissimilarity of these relativistic finite chiral chemical potential and temperature terms to Eqs. \eqref{NewtonG}, \eqref{A0}, and \eqref{A0mu}.
 
%\date{\today}

\end{document}